\documentstyle[aps,prl,multicol,epsfig]{revtex}

\begin{document}

\title{New Class of Eigenstates in Generic Hamiltonian Systems}
\author{R.~Ketzmerick, L.~Hufnagel, F.~Steinbach, M.~Weiss \\
Max-Planck-Institut f\"ur Str\"omungsforschung\\
and Institut f\"ur Nichtlineare Dynamik der Universit\"at G\"ottingen,\\
Bunsenstr.~10, 37073 G\"ottingen, Germany}
\date{\today}
\maketitle

\begin{abstract}
In mixed systems, besides regular and chaotic states, there are states 
supported by the chaotic region mainly living in the
vicinity of the hierarchy of regular islands. We show that the fraction 
of these {\it hierarchical states} scales as~$\hbar^{-\alpha}$ and relate
the exponent~$\alpha=1-1/\gamma$ to the decay of the classical staying 
probability~$P(t)\sim t^{-\gamma}$. This is numerically
confirmed for the kicked rotor by studying the influence of hierarchical states
on eigenfunction and level statistics.
\end{abstract}

\begin{multicols}{2}
Typical Hamiltonian systems are neither integrable nor ergodic~\cite{MM74} but have a {\it mixed}
phase space, where regular and chaotic regions coexist. The regular regions are organized in a
hierarchical way~\cite{lichtenberg} (see, e.g., Fig.~2a) and chaotic dynamics is clearly distinct from
the dynamics of fully chaotic systems. In particular, chaotic trajectories are trapped in the 
vicinity of the hierarchy of regular islands. The most prominent quantity reflecting this, is the
probability~$P(t)$ to be trapped longer than a time~$t$, which decays as~\cite{powerlaw}
\begin{equation}\label{poft}
P(t)\sim t^{-\gamma},\quad \gamma>1\,\, ,
\end{equation}
in contrast to the typically exponential decay in fully chaotic systems.
While the power-law decay is universal, the exponent~$\gamma$ is system and parameter dependent.
The origin of the algebraic decay are partial transport barriers~\cite{HCM85}, 
e.g., Cantori, leading to a hierarchical structure of the chaotic region~\cite{kay}.
Quantum mechanically the classical algebraic decay of~$P(t)$ is mimicked at most until
Heisenberg time~\cite{maspero}.

Even after two decades of studying quantum chaos, the search for quantum signatures of this
universal power-law trapping is still in its infancy: In fact,
only conductance fluctuations of open systems have been investigated so far. It was semiclassically
derived that these fluctuations should have a fractal dimension~$D=2-\gamma/2$~\cite{roland}, which
was confirmed in gold nanowires~\cite{hegger}, semiconductor nanostructures~\cite{sach}, 
and numerics~\cite{italo}. Quite recently, a second type of conductance fluctuations in 
mixed systems has been discovered numerically~\cite{bodo,wir}, namely isolated resonances. 
There the classical exponent~$\gamma$ seems to appear in the scaling of the variance of conductance 
increments, surprisingly, on scales {\it below} the mean level spacing, what is not understood so far. 
Thus, even for this subject there is a lack of basic understanding.

\begin{figure}
\begin{center}
\epsfig{figure=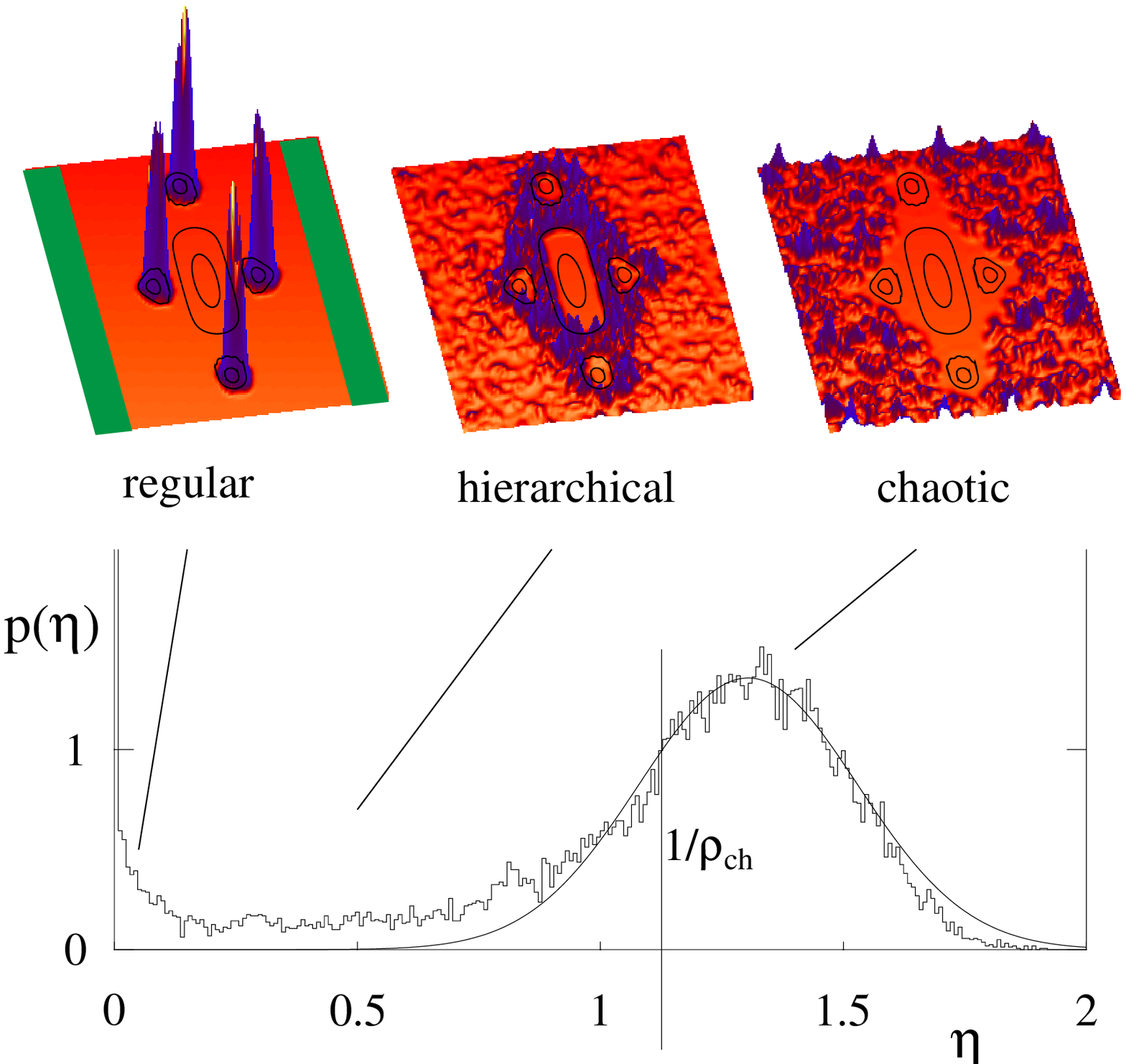,width=8cm}\hfill
\end{center}
\noindent
{\footnotesize {\bf FIG.~1.} Husimi representation of a regular,
hierarchical, and chaotic eigenstate of the kicked rotor ($K=2.5,\hbar=2\pi/1000$) with 
solid lines showing KAM-tori of the classical phase space. The lower part shows the 
distribution~$p(\eta)$ of the average density~$\eta$ an eigenstate has in the chaotic region
far away from the islands (green shaded area corresponding to 25\% of sites in $q$-representation). 
The mean of the Gaussian fit determines~$1/f_{\rm ch}$, which is clearly distinct from~$1/\rho_{\rm ch}$.}
\end{figure}

In this paper we present consequences of the classical~$P(t)\sim t^{-\gamma}$ in closed quantum
systems, namely the appearance of a new class of eigenstates. Different from the well studied
regular and chaotic states, they are supported by the chaotic region, but predominantly live in 
the vicinity of the regular islands~\cite{radons}. As for decreasing~$\hbar$ they move deeper into the
hierarchical structure of the chaotic region, we like to call them {\it hierarchical} states.
We show that the fraction~$f_{\rm hier}$ of such states scales as
\begin{equation}\label{scaling}
f_{\rm hier}\sim \hbar^{1-1/\gamma}\,\, ,
\end{equation}
which is confirmed numerically for the kicked rotor by studying their influence on eigenfunction 
and level statistics. We deduce relation~(\ref{scaling}) by combining the finite resolution of quantum 
mechanics for a given~$\hbar$ with the simplest model~\cite{HCM85} 
describing classical transport in a mixed phase 
space. We are encouraged by the success of this approach, and think that it will help
in understanding the quantum signatures of a mixed phase space.

We motivate and numerically verify our results using the well known kicked rotor, which 
is a paradigm for a generic Hamiltonian system~\cite{chir} and has a time
evolution described by the map:
\begin{eqnarray}\label{standardmap}
q_{n+1}&=&q_n+p_n \nonumber\\ p_{n+1}&=&p_n+K\sin q_{n+1} \,\, .
\end{eqnarray}
Its quantum properties are determined by the time evolution operator for one period~\cite{CCFI79,chang}
\begin{equation}\label{u}
{\cal U}=\exp\left(-\frac{i}{\hbar} K\cos q\right)
\exp\left(-\frac{i}{\hbar}\frac{p^2}{2}\right) \,\, .
\end{equation}
Fig.~2a shows a typical classical phase space, which is governed by one big chaotic invariant 
set with fractional phase space volume~$\rho_{\rm ch}$ and islands of regular motion 
of total fraction~$\rho_{\rm reg}$~\cite{bem1} arranged in a hierarchical way. 
Quantum mechanically, however, we find {\it three} types of eigenstates of~$\cal U$ (Fig.~1)~: 
There are 'regular' states living on KAM-tori of the regular islands and there are 'chaotic' 
states extending across most of the chaotic region as first described by Percival~\cite{percival}. 
In addition, there are eigenstates supported by the chaotic region, that are of a third type. 
They predominantly live in the vicinity of the regular islands with only a small
contribution in the main part of the chaotic sea.
These states are separated from the main chaotic sea by partial barriers of the classical 
phase space, e.g., Cantori or stable and unstable manifolds~\cite{kay}.
For decreasing~$\hbar$ these states live deeper in the hierarchy of classical 
phase space and thus we like to call them {\it hierarchical} states. Denoting the fraction of the three
types of states by~$f_{\rm reg}, f_{\rm ch}$ and~$f_{\rm hier}$, respectively, one has 
\begin{equation}\label{fsumme}
f_{\rm reg} + f_{\rm ch} + f_{\rm hier}=1=\rho_{\rm reg}+\rho_{\rm ch}\,\, .
\end{equation}
Obviously for any given~$\hbar$ not all regular islands are resolved and 
therefore~$f_{\rm reg}(\hbar)<\rho_{\rm reg}$ holds. 
We determined the island volumes, revealing that the $\hbar$-dependence of~$f_{\rm reg}$
is negligible compared to the $\hbar$-dependence of~$f_{\rm hier}$ and~$f_{\rm ch}$. We thus use
$f_{\rm reg}=\rho_{\rm reg}$ in the following and Eq.~(\ref{fsumme}) reduces to
\begin{equation}\label{redf}
f_{\rm ch} + f_{\rm hier}=\rho_{\rm ch}\,\, .
\end{equation}
We want to focus on the fraction~$f_{\rm hier}$ of hierarchical states, which we now determine
from their influence on eigenfunction and level statistics. 

\begin{figure}
\begin{center}
\epsfig{figure=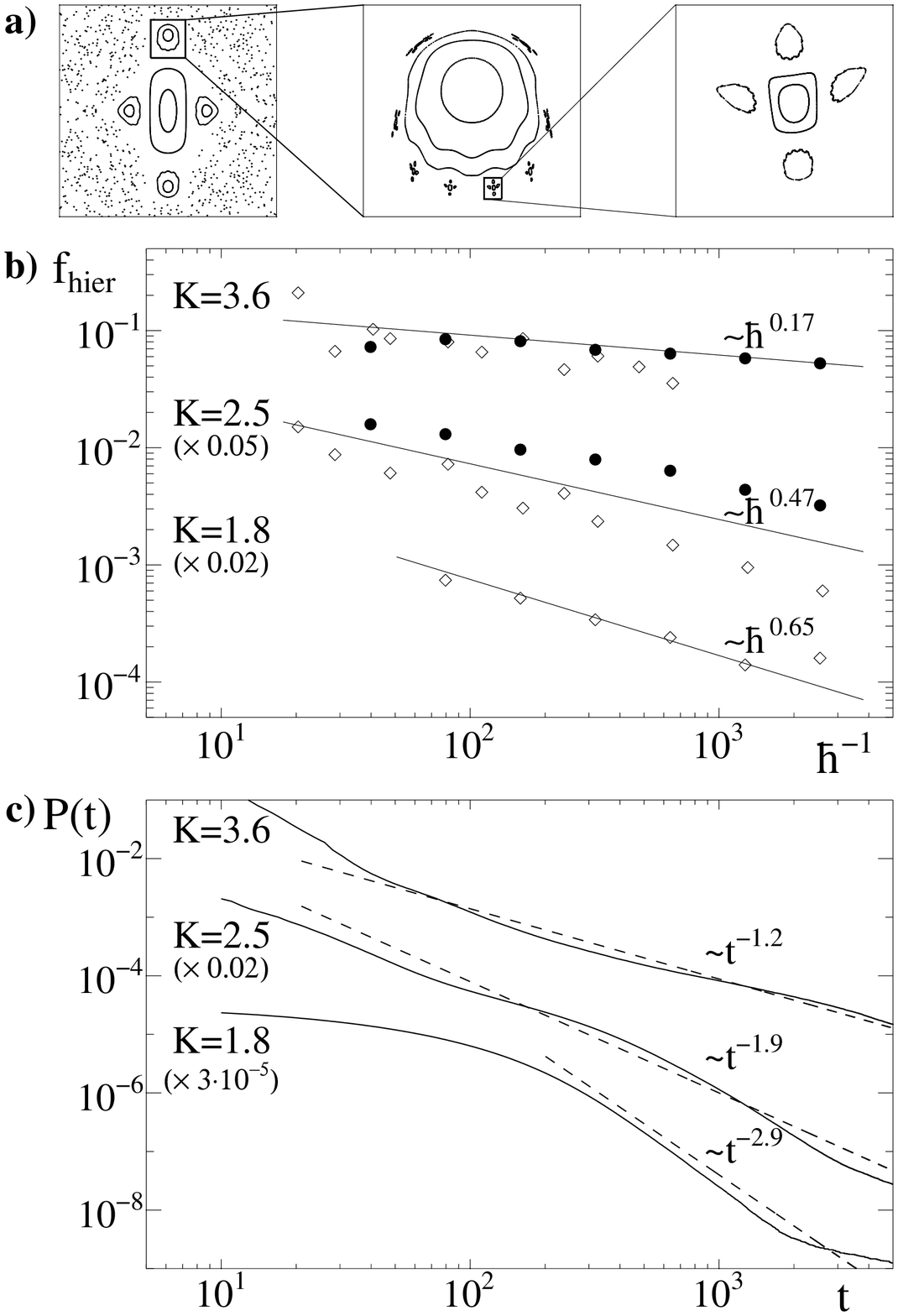,width=8cm}\hfill
\end{center}
\noindent
{\footnotesize {\bf FIG.~2.} {\bf a)} Successive magnifications of the phase space hierarchy 
of the kicked rotor ($K=2.5$). 
{\bf b)} The fraction of hierarchical states~$f_{\rm hier}$ as a function of~$1/\hbar$ for
$K=3.6,2.5,1.8$ as determined from eigenfunction (dots) and level statistics (diamonds). The
solid lines show the predicted power laws according to Eq.~(\ref{scaling}) and Fig.~2c.
Data for~$K=1.8$ are extracted from table~2 in Ref.~\cite{pr}.
{\bf c)} Classical staying probability~$P(t)\sim t^{-\gamma}$ providing the values of~$\gamma$ used
in b). The reason for the different fitting range for~$K=1.8$ is the longer lasting initial exponential
decay.}
\end{figure}

As a first quantity we determine~$\eta$, the density of an eigenstate averaged over a large area
in the chaotic region, far away from the islands. As long as this averaging area is large compared
to~$\hbar$ its specific choice and the used representation (e.g., Husimi) does not affect the following 
analysis. Normalization for~$\eta$ is chosen such that, an eigenstate uniformly covering the entire phase 
space would give~$\eta=1$. The distribution of~$\eta$ (Fig.~1) shows a peak for small~$\eta$ 
stemming from the regular states, intermediate values from hierarchical states and a Gaussian-like 
peak that we ascribe to the chaotic states. Assuming that the fraction~$f_{\rm ch}$ of chaotic 
states extends on average uniformly (neglecting scar effects~\cite{heller}) across a 
fraction~$f_{\rm ch}\le\rho_{\rm ch}$ of phase space, we fit a Gaussian with 
area~$f_{\rm ch}$ and mean~$1/f_{\rm ch}$ to the latter part of the $\eta$-distribution. 
We find that the mean~$1/f_{\rm ch}$ is clearly distinct from~$1/\rho_{\rm ch}$ (Fig.~1), 
the value one would obtain
if there were eigenstates uniformly covering the {\it entire} chaotic region.
The fit determines the fraction $f_{\rm hier}=\rho_{\rm ch}-f_{\rm ch}$ of hierarchical 
eigenstates, which is shown as a function of~$\hbar$ in Fig.~2b.

Secondly, we want to quantify the influence of hierarchical states on level statistics for mixed 
systems. To this end we extend the approach of Berry and Robnik~\cite{br}, which assumes a 
random superposition of a regular (Poissonian) spectrum with measure~$\rho_{\rm BR}$ 
and a chaotic (GOE) spectrum with measure~$1-\rho_{\rm BR}$~: We take into account the 
hierarchical states and their spectrum.
As most of them couple weakly to regular as well as chaotic states, their spectrum is superimposed 
independently to the rest of the spectrum. Typically, $f_{\rm hier}$ is smaller 
than~$f_{\rm ch}+f_{\rm reg}$ and therefore to first order these levels are randomly placed in 
the spectrum. Thus, effectively one can use the original Berry-Robnik approach, but one has to 
interpret the parameter~$\rho_{\rm BR}$ as the sum of regular and hierarchical fraction of states,
$\rho_{\rm BR}=f_{\rm reg}+f_{\rm hier}$. We determine~$\rho_{\rm BR}$ by fitting the nearest-neighbor 
level-spacing distribution, using the cumulative distribution as well as 
the so-called U-function~\cite{pr}. The obtained values of~$f_{\rm hier}$ from both fitting 
procedures agree within the error bars and their average as a function of~$\hbar$ is shown in Fig.~2b.

The two methods for determining~$f_{\rm hier}$, eigenfunction and level statistics, rely on several
assumptions, and one may not expect identical absolute values. Both methods, however, show a 
clear power-law decay according to relation~(\ref{scaling}) 
for the fraction of hierarchical eigenstates as can be seen in Fig.~2b for three values of the
kicking strength~$K$, where~$\gamma$ is extracted from the decay 
of the corresponding~$P(t)$ shown in Fig.~2c.

We now want to derive Eq.~(\ref{scaling}). Let us remind that 
the classical power-law trapping~$P(t)\sim t^{-\gamma}$ originates from partial transport 
barriers, e.g., Cantori, arranged in a hierarchical way in the chaotic part of phase space
around islands of regular motion~\cite{kay}. 
The simplest model~\cite{HCM85} yielding~$P(t)\sim t^{-\gamma}$
describes this hierarchy of the chaotic sea 
as a chain of volumes~$\Omega_n\sim\omega^n$ ($\omega<1,\,n=0,1,\ldots$) 
connected by a flux~$\Phi_{n,n+1}\sim\varphi^n$ ($\varphi<1$) as shown in Fig.~3a. 
Then~$\gamma=1/(1-\ln\omega/\ln\varphi)$ holds~\cite{HCM85}. 
Quantum mechanically two neighboring volumes~$n$ and~$n+1$ are strongly coupled, as long as
$\Phi_{n,n+1}>\hbar$, while they are weakly coupled in the opposite case~\cite{bohigas}.
This determines a critical flux~$\Phi_{n^*,n^*+1}=\hbar$ with~$n^*\sim \ln\hbar/\ln\varphi$.
The volumes~$\Omega_n$ with~$n<n^*$ correspond to the main part of the chaotic sea and support
the chaotic eigenstates. Regions with~$n>n^*$ correspond to the hierarchical part of the chaotic sea
supporting the hierarchical states. Summation of their volumes, finally, yields the fraction
\begin{figure}
\begin{center}
\epsfig{figure=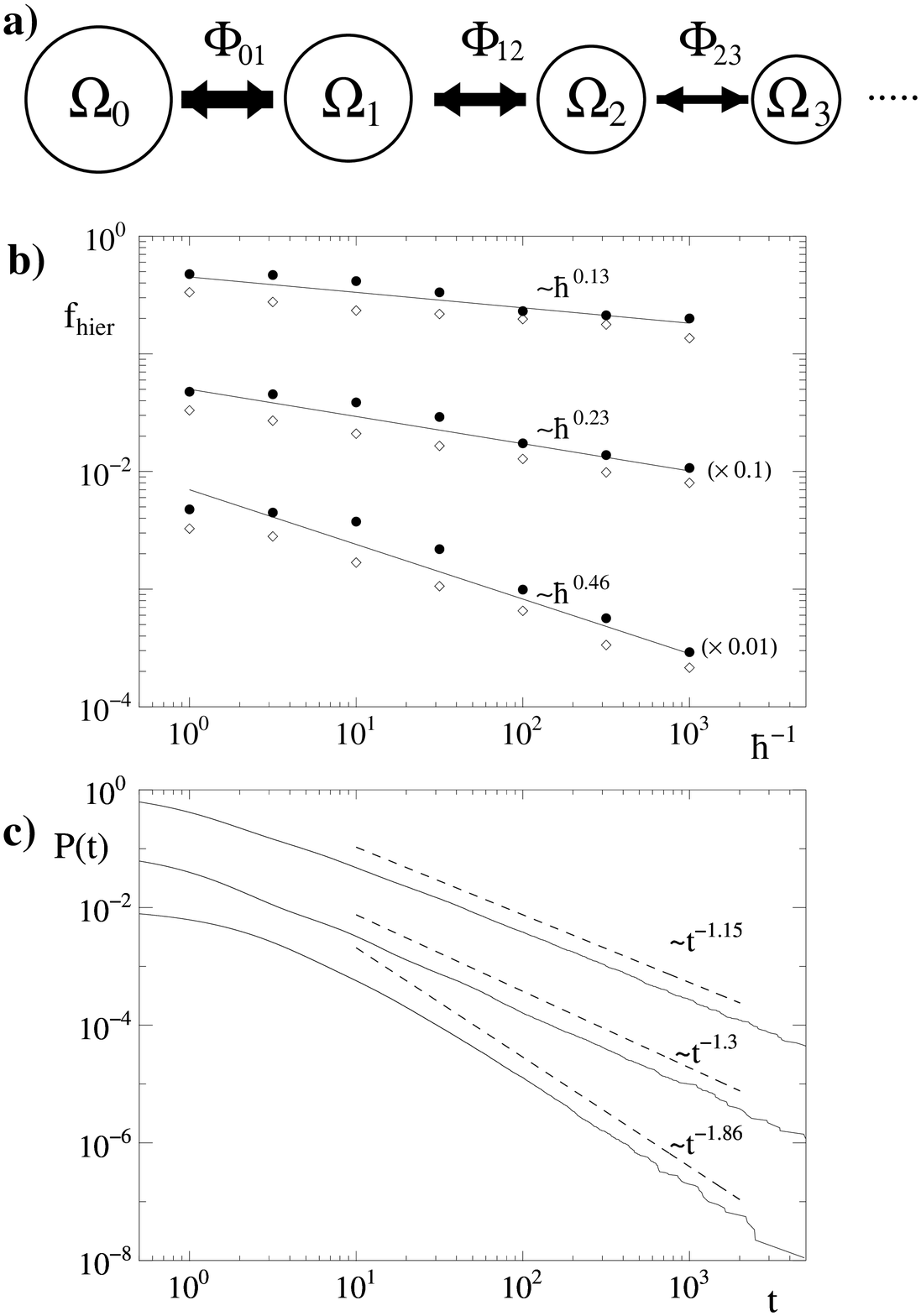,width=8cm}\hfill
\end{center}
\noindent
{\footnotesize {\bf FIG.~3.} {\bf a)} Sketch of the chain model with scaling volumes and fluxes.
{\bf b)} The fraction of hierarchical states~$f_{\rm hier}$ for three random matrix ensembles
($\gamma=1.15,1.3,1.86$) as a function of~$1/\hbar$ 
as determined from eigenfunction (dots) and level statistics (diamonds). 
The solid lines show the predicted power laws according to Eq.~(\ref{scaling}).
{\bf c)} Classical staying probability~$P(t)\sim t^{-\gamma}$ confirming the values of~$\gamma$ used
in b).}
\end{figure}
\begin{equation}\label{rmtskal}
f_{\rm hier}\sim\sum_{n>n^*} \Omega_n\sim\omega^{n^*}\sim\hbar^{1-1/\gamma}\,\, .
\end{equation}
of hierarchical states. From~$\gamma>1$ it follows, that in the limit~$\hbar\to 0$ this fraction
tends to zero, while the total number of hierarchical states goes to infinity.
In case of a system with~$d\neq 2$ pairs of conjugate variables,
one needs to replace~$\hbar$ by~$\hbar^{d-1}$ in the above derivation.

In order to check relation~(\ref{rmtskal}), we generated a random matrix model for the
classical chain following the approach of Ref.~\cite{bohigas}. For each ensemble 
of these random matrices we fixed~$\omega,\varphi$ and therefore also~$\gamma$
and then varied~$\hbar$. We repeated the above eigenfunction and level statistics
and extracted~$f_{\rm hier}$ ($f_{\rm reg}=\rho_{\rm reg}=0$ in this model).
The results for various~$\gamma$ are shown in Fig.~3b. The numerical data 
agree very well with the expected power laws over three orders of magnitude in~$\hbar$ and 
therefore confirm relations~(\ref{rmtskal}) and~(\ref{scaling}).

In conclusion, we present numerical evidence that besides chaotic and regular eigenfunctions 
there exists a third class, namely hierarchical eigenstates.
Their fraction is determined from eigenfunction and level statistics and shows a power law
as a function of~$\hbar$. We explain the origin of the 
power law and relate its exponent to the well known power-law trapping of mixed systems.

We like to thank S.~Tomsovic and L.~Kaplan for helpful discussions.

\end{multicols}


\begin{thebibliography}{1}
\bibitem{bem1} We determined~$\rho_{\rm ch}$ ($\pm10^{-3}$) by measuring the phase space volume 
explored by a chaotic trajectory. For the present purpose we may include into~$\rho_{\rm reg}$
the tiny chaotic sets within the islands of regular motion.
\end{thebibliography}
\end{document}